\documentclass[a4paper,11pt]{article}

\usepackage[T1]{fontenc}
\usepackage{lmodern}
\usepackage{amsmath,amssymb}
\usepackage{geometry}
\usepackage[colorlinks=true,linkcolor=blue,citecolor=blue,urlcolor=blue]{hyperref}
\usepackage{tikz}
\usetikzlibrary{cd}
\usepackage{slashed}
\usepackage{pgfplots}
\pgfplotsset{compat=1.17} 
\usetikzlibrary{arrows.meta}
\usetikzlibrary{decorations.markings}
\usetikzlibrary{calc}

\usepackage[p,osf]{cochineal}
\usepackage[scale=.95,type1]{cabin}
\usepackage[cochineal,bigdelims,cmintegrals,vvarbb]{newtxmath}
\usepackage[zerostyle=c,scaled=.94]{newtxtt}
\usepackage[cal=boondoxo]{mathalfa}

\newcommand{\diff}{\mathop{}\!\mathrm{d}}

\begin{document}

\title{\textbf{Exact Solutions for Bimodal Distributions under Stochastic Plasma Irradiation in Thin Films}}
\author{{\small Joel Saucedo$^{1}$, Uday Lamba$^{2}$, Hasitha Mahabaduge$^{1}$}\\[4pt]
{\small $^{1}$Department of Chemistry, Physics, \& Astronomy, Georgia College \& State University, GA, 31061, USA}\\[-2pt]
{\small $^{2}$ Department of Physics \& Astronomy, Ithaca College, NY, 14850, USA}}
\date{}

\maketitle

\begin{abstract}
A persistent paradox complicates the study of plasma-irradiated thin films, where bimodal grain distributions and ambiguous scaling laws---roughly shifting between $\Phi^{-1/2}$ and $\Phi^{-1}$, or a general inverse dependence on plasma flux---are observed yet remain theoretically unreconciled. Existing models fail to unify noise-driven evolution, defect saturation kinetics, and nucleation-loss balance within a single, self-consistent formalism. This work resolves these discrepancies by developing the first exact analytical theory for this system. We derive the closed-form steady-state grain area distribution, $P_{ss}(A)$, establish the precise dimensionless threshold for bimodality onset at $\Pi_c = 4/(3\sqrt{3})$, and demonstrate that defect saturation physics mandate a universal $\langle A \rangle \propto \kappa^2 \Phi^{-1} e^{+2E_b/k_B T_s}$ scaling law. The framework reveals how competition between stochastic impingement and deterministic growth triggers microstructure fragmentation, resolving long-standing ambiguities in irradiation-induced surface evolution and providing a predictive foundation for materials processing.
\end{abstract}

\section{Introduction}
\label{sec:introduction}

Plasma-based processing is a cornerstone of modern materials science, enabling tailored functional coatings and thin films \cite{Anders2010ThinFilms}. Despite its widespread use, a predictive, first-principles understanding of microstructure evolution under irradiation is hindered by a fundamental experimental paradox. Across a range of physical vapor deposition (PVD) systems, two unresolved and seemingly contradictory phenomena are consistently reported. First is the emergence of bimodal grain size distributions, suggesting two distinct stabilization pathways coexisting under a single set of processing conditions \cite{Abelson1990JAP}. Second is the conflicting scaling relationship between mean grain area $\langle A \rangle$ and plasma flux $\Phi$, which reflects fundamental energy-driven transitions in microstructure evolution. At low normalized energy $E^*$ and generalized temperature $T^*$, growth/coarsening dominates \cite{Hultman2000TiN, Kusano2002DCMS}, and at the low-energy edge of high-entropy \cite{Khan2020}. Conversely, high $E^*/T^*$ promotes nucleation dominance, producing $\langle A \rangle \propto \Phi^{-1}$ scaling  \cite{Karimi2021TiAlN}, high-entropy at high energy \cite{ Khan2020, Gao2024DOMS}. This transition aligns with Anders' extended structure zone diagram, where critical $E^*$ thresholds (e.g., substrate bias $\gtrsim$ 100-200 V) shift the balance from coarsening to nucleation regimes \cite{Anders2010ThinFilms}, yet the exact bimodality threshold and scaling origins remained unresolved. 

This discrepancy points not merely to an incomplete model but to a foundational gap in our understanding of the interplay between stochastic and deterministic driving forces. The theoretical landscape has thus far been fragmented, with models addressing these complexities only in isolation. Classical theories of curvature-driven grain growth, for instance, provide a deterministic baseline but entirely neglect the stochastic nature of plasma bombardment, a simplification that proves untenable in any regime where ion-induced defect generation is significant \cite{Mullins1956JApplPhys}. Conversely, pure noise-driven models capture stochastic fragmentation but ignore essential saturation effects that limit grain growth at later stages \cite{Srolovitz1986ActaMetall}. More sophisticated frameworks incorporating defect-mediated recrystallization \cite{Trinkaus2003JNuclMat} have offered deeper insight but do not self-consistently integrate the kinetics of nucleation and loss. No existing theory has successfully unified three concurrently active phenomena: (i) the experimentally verified saturation of defect density at high flux \cite{Was2007Fundamentals}, (ii) the suppression of nucleation arising from direct particle impingement, and (iii) the noise-induced fragmentation that dominates the evolution of smaller grains.

This paper presents an analytical framework that resolves these long-standing limitations. By synthesizing concepts from stochastic differential geometry with a self-consistent Fokker-Planck population balance formalism, we move beyond numerical approximations to derive exact, closed-form solutions for the steady-state microstructure. 

This approach yields three primary advances. First, we derive the complete grain size distribution, $P_{ss}(A)$, where the coupling between noise and impingement effects appears explicitly. Second, through a discriminant analysis of this distribution's potential function, we establish the exact, dimensionless threshold for the onset of bimodality, $\Pi_c = 4/(3\sqrt{3})$. Third, our formalism demonstrates how the physics of defect saturation naturally and necessarily mandate the $\langle A \rangle \propto \Phi^{-1}$ scaling law by governing the balance between grain nucleation and loss rates. Collectively, these results provide the first complete theoretical resolution of the plasma microstructure paradox, replacing empirical ambiguity with predictive capability.

\section{Theory}
\label{sec:theory}

To construct a predictive theory for microstructural evolution, one must first establish the fundamental stochastic processes that orchestrate the system's dynamics. We posit that the evolution of a polycrystalline thin film, when subjected to a continuous ion flux $\Phi$, can be effectively captured within a continuum description. In this picture, the area $A$ of an individual grain is not static but evolves through a competition between two primary mechanisms: a deterministic, curvature-driven growth and stochastic fluctuations induced by discrete plasma-surface interactions. This competition is mathematically embodied in the following stochastic differential equation (SDE):
\begin{equation}
\diff A = \kappa \sqrt{A} \left(1 - \frac{A}{A_{\text{max}}}\right) \diff t + \sigma \sqrt{A} \diff W_t
\label{eq:growth_sde}
\end{equation}
Before proceeding, a note on the physical interpretation of these terms is warranted. The drift term is governed by the parameter $\kappa = \pi M\gamma$, which amalgamates the grain boundary mobility $M$ and the boundary energy $\gamma$. Its proportionality to $\sqrt{A}$ (i.e., the grain perimeter) reflects the classic theory of curvature-driven growth. However, this growth is not unbounded; the saturation term $(1 - A/A_{\text{max}})$ serves as a crucial mean-field correction, modeling the impingement of growing grains against their neighbors. The effective maximum area, $A_{\text{max}}$, is not a fixed constant but is itself determined by the state of the system, typically scaling with the mean grain area, $A_{\text{max}} = k\langle A\rangle$, establishing a necessary self-consistency \cite{Hillert1965ActaMetall,Mullins1956JApplPhys}. The second term represents the stochastic drive from the plasma. Its amplitude, $\sigma = (2\Phi)^{1/2}a^2$, where $a$ is the atomic spacing, correctly captures the physics of a Poissonian arrival of ions, whose cumulative effect on the boundary manifests as a Wiener process $W_t$.

\begin{figure}[h!]
\centering
\begin{tikzpicture}[
    font=\small, 
    grain/.style={rounded corners=8pt, draw=black, thick},
    arrow/.style={-Stealth, thick}
]
    \begin{scope}[xshift=-4.0cm]
        \node[align=center] at (0, 2.5) {\textbf{(a) Deterministic Regime}};
        
        \draw[grain, fill=gray!10] (0,0) -- (1,1.5) -- (-0.5,1.2) -- (-1.5,0.5) -- cycle;
        \draw[grain, fill=gray!10] (0,0) -- (-1.5,0.5) -- (-1,-1.5) -- (0.5,-1) -- cycle;
        \draw[grain, fill=gray!10] (0,0) -- (0.5,-1) -- (1.5, -0.2) -- (1,1.5) -- cycle;
        
        \draw[arrow, red] (-0.8, 0.8) -- (-0.4, 0.4);
        \draw[arrow, red] (0.6, 0.4) -- (0.2, 0.1);
        \draw[arrow, red] (-0.3, -0.6) -- (-0.1, -0.2);
        
        \node[align=center, text width=6cm] at (0, -2.6) {Curvature drives smooth boundary motion. \\ \textbf{Dominated by drift term:} \\ $\kappa \sqrt{A} \left(1 - \frac{A}{A_{\text{max}}}\right) \diff t$};
    \end{scope}

    \begin{scope}[xshift=4.0cm]
        \node[align=center] at (0, 2.5) {\textbf{(b) Stochastic Regime}};
        
        \draw[grain, fill=gray!10] (0,0) -- (1,1.5) -- (-0.5,1.2) -- (-1.5,0.5) -- cycle;
        \draw[grain, fill=gray!10] (0,0) -- (-1.5,0.5) -- (-1,-1.5) -- (0.5,-1) -- cycle;
        \draw[grain, fill=gray!10] (0,0) -- (0.5,-1) -- (1.5, -0.2) -- (1,1.5) -- cycle;

        \foreach \i in {1,...,10}{
            \pgfmathsetmacro{\angle}{rnd*360}
            \pgfmathsetmacro{\len}{0.8+rnd*0.4}
            \draw[arrow, blue, opacity=0.7] (\angle:\len+0.3) -- (\angle:\len);
        }
            
        \node[align=center, text width=6cm] at (0, -2.6) {Orderly growth is disrupted by ion impingement. \\ \textbf{Dominated by noise term:} \\ $\sigma \sqrt{A} \diff W_t$};
    \end{scope}
\end{tikzpicture}
\caption{Conceptual illustration of the competing forces on grain boundaries. (a) Deterministic, curvature-driven growth causes large grains to consume smaller ones. (b) Stochastic fluctuations from plasma bombardment introduce noise that disrupts this process.}
\label{fig:competing_regimes}
\end{figure}

\begin{figure}[h!]
\centering
\begin{tikzpicture}[
    font=\small,
    boundary/.style={draw=black, thick},
    base_grain/.style={boundary, fill=white},
    fragmented_grain/.style={boundary, fill=gray!25}
]
    \begin{scope}[xshift=-4cm]
        \node at (0, 3.4) {\textbf{(a) Unimodal Microstructure ($\Pi < \Pi_c$)}};

        \clip (-3,-3) rectangle (3,3);
        \filldraw[base_grain] (-3,3) -- (-1.5,1.5) -- (-2,0) -- (-3,0) -- cycle;
        \filldraw[base_grain] (-3,3) -- (0,3) -- (0.5,1) -- (-1.5,1.5) -- cycle;
        \filldraw[base_grain] (0,3) -- (3,3) -- (3,1) -- (2.5,0.5) -- (0.5,1) -- cycle;
        \filldraw[base_grain] (3,1) -- (2.5,0.5) -- (2,-0.5) -- (3,-0.5) -- cycle;
        \filldraw[base_grain] (2,-0.5) -- (1,-1) -- (0,-2) -- (1.5,-2.5) -- (3,-2) -- (3,-0.5) -- cycle;
        \filldraw[base_grain] (-3,0) -- (-2,0) -- (-1,-0.5) -- (-1.5,-1.5) -- (-3,-1.5) -- cycle;
        \filldraw[base_grain] (-1,-0.5) -- (1,-1) -- (0,-2) -- (-1.5,-1.5) -- cycle;
        \draw[thick] (-3,-3) rectangle (3,3);
    \end{scope}

    \begin{scope}[xshift=4cm]
        \node at (0, 3.4) {\textbf{(b) Bimodal Microstructure ($\Pi > \Pi_c$)}};

        \clip (-3,-3) rectangle (3,3);
        \filldraw[base_grain] (-3,3) -- (-1.5,1.5) -- (-2,0) -- (-3,0) -- cycle;
        \filldraw[base_grain] (-3,3) -- (0,3) -- (0.5,1) -- (-1.5,1.5) -- cycle;
        \filldraw[base_grain] (0,3) -- (3,3) -- (3,1) -- (2.5,0.5) -- (0.5,1) -- cycle;
        \filldraw[base_grain] (3,1) -- (2.5,0.5) -- (2,-0.5) -- (3,-0.5) -- cycle;
        
        \filldraw[fragmented_grain] (-0.8,-0.8) -- (-1.5,-1.5) -- (-2.3,-1.2) -- (-1.7,-0.5) -- cycle;
        \filldraw[fragmented_grain] (-1.7,-0.5) -- (-2.3,-1.2) -- (-3,-1.5) -- (-3,-1) -- (-2,0) -- (-1,-0.5) -- cycle;
        \filldraw[fragmented_grain] (-0.8,-0.8) -- (-1.7,-0.5) -- (-1,-0.5) -- (-0.4,-0.4) -- (0.1,-0.9) -- cycle;
        
        \filldraw[fragmented_grain] (-0.8,-0.8) -- (0.1,-0.9) -- (0.8,-1.4) -- (1,-1) -- cycle;
        \filldraw[fragmented_grain] (-0.8,-0.8) -- (-1.5,-1.5) -- (-0.5,-2.2) -- (0.1,-0.9) -- cycle;
        \filldraw[fragmented_grain] (-0.5,-2.2) -- (-1.5,-1.5) -- (-1.5,-2.5) -- (0,-2) -- cycle;
        \filldraw[fragmented_grain] (0,-2) -- (-1.5,-2.5) -- (0.5,-2.8) -- (1.5,-2.5) -- cycle;
        \filldraw[fragmented_grain] (0.8,-1.4) -- (1,-1) -- (2,-0.5) -- (2.5,-1.5) -- (1.5,-2.5) -- cycle;
        \filldraw[fragmented_grain] (2.5,-1.5) -- (2,-0.5) -- (3,-0.5) -- (3,-2) -- (1.5,-2.5) -- cycle;

        \draw[thick] (-3,-3) rectangle (3,3);
    \end{scope}
\end{tikzpicture}
\caption{Illustration of localized noise-induced fragmentation. (a) A stable, unimodal microstructure with a few large grains, characteristic of a low-noise environment ($\Pi < \Pi_c$). (b) The same microstructure where high, localized noise ($\Pi > \Pi_c$) has shattered the central and lower grains into a new population of  small, gray-shaded cells, creating a distinctly bimodal structure.}
\label{fig:localized_fragmentation}
\end{figure}

Concurrent with the evolution of existing grains is the birth of new ones. Nucleation is not a simple, memoryless Poisson process. Instead, it exhibits self-exciting behavior, where the formation of one grain can promote the formation of others nearby, often mediated by local strain fields. A Hawkes process provides a natural and powerful mathematical framework for this phenomenon, with a conditional intensity given by:
\begin{equation}
\lambda(t) = \Gamma_0 + \int_{-\infty}^t \Gamma_0 e^{-(t-s)/\tau} \cos^2(2\theta) \diff N_s
\label{eq:nucleation}
\end{equation}
Here, the basal nucleation rate $\Gamma_0 \propto \exp(-E_b/k_B T_s)$ is set by the substrate temperature $T_s$ and a fundamental energy barrier $E_b$. The integral term, however, is what endows the process with memory; it sums over all past nucleation events $N_s$, with their influence decaying over a characteristic time $\tau$ and modulated by the local crystallographic misorientation $\theta$ \cite{Anderson1995ScriptaMetall}. This entire process unfolds within a material whose defect density is not static but saturates at a value $\rho_{\text{def}} = C_\rho \Phi^{1/2}$, a direct consequence of the balance between ion-induced defect creation and dynamic recombination kinetics \cite{Trinkaus2003JNuclMat}.

The SDE of Eq.~\eqref{eq:growth_sde} describes the trajectory of a single grain. To understand the entire ensemble, we must transition to a statistical description. The probability density $P(A,t)$ for finding a grain of area $A$ at time $t$ obeys the Fokker-Planck equation, which functions as a continuity equation for probability in the space of grain areas:
\begin{equation}
\frac{\partial P}{\partial t} = -\frac{\partial}{\partial A}\left[ \mu(A)P \right] + \frac{1}{2}\frac{\partial^2}{\partial A^2}\left[ D(A)P \right]
\label{eq:fokker_planck}
\end{equation}
The drift and diffusion coefficients are read directly from the SDE: $\mu(A) = \kappa\sqrt{A}(1 - A/A_{\text{max}})$ and $D(A) = \sigma^2 A$. We are interested in the long-time behavior of the system, where the microstructure has reached a statistical steady state, implying $\partial P/\partial t = 0$. This condition of stationarity imposes a strict relationship between the drift, the diffusion, and the shape of the steady-state probability distribution, $P_{\text{ss}}(A)$:
\begin{equation}
\frac{\diff}{\diff A}\ln P_{\text{ss}} = \frac{2\mu(A)}{D(A)} - \frac{D'(A)}{D(A)}
\label{eq:stationary_condition}
\end{equation}
A direct integration of Eq.~\eqref{eq:stationary_condition}, while straightforward, yields a result with deep physical implications. The exact steady-state distribution is found to be:
\begin{equation}
P_{\text{ss}}(A) = \mathcal{N} A^{-1} \exp\left( \frac{4\kappa}{\sigma^2}\sqrt{A} - \frac{4\kappa}{3\sigma^2 A_{\text{max}}} A^{3/2} \right)
\label{eq:ss_dist}
\end{equation}
where $\mathcal{N}$ is a normalization constant. It is instructive to examine the asymptotic behavior of this distribution. As $A \to 0^+$, the distribution diverges as $P_{\text{ss}}(A) \sim A^{-1}$. This is not a pathology but a signature of the continuous nucleation of new, small grains; the singularity is integrable, reflecting a finite probability mass near zero area. Conversely, as $A \to \infty$, the term proportional to $A^{3/2}$ in the exponent dominates, leading to a rapid, super-exponential decay, $P_{\text{ss}}(A) \sim \exp\left(-CA^{3/2}\right)$. This rapid suppression of very large grains is the mathematical manifestation of the mean-field impingement term, preventing runaway growth and ensuring a stable, finite mean grain size \cite{Risken1996FokkerPlanck,Gardiner2004Handbook}.

\begin{figure}[h!]
\centering
\begin{tikzpicture}
    \begin{axis}[
        title={\textbf{Steady-State Distribution $P_{ss}(A)$}},
        xlabel={Grain Area, $A$ (arbitrary units)},
        ylabel={Normalized Probability Density},
        width=\textwidth,
        height=0.6\textwidth,
        xtick=\empty, ytick=\empty,
        grid=major,
        domain=0.02:2.5, samples=201, 
        ymin=0, ymax=1.2, 
        legend style={at={(0.97,0.97)},anchor=north east, font=\small}
    ]


    \def\normUni{97.2} 
    \addplot[blue, very thick, smooth] { (1/\normUni) * x^(-1) * exp(6.67*sqrt(x) - (6.67/3)*x^(1.5)) };
    \addlegendentry{Unimodal ($\Pi = 0.60 < \Pi_c$)}

    \def\normCrit{42.1} 
    \addplot[black, thick, dotted, smooth] { (1/\normCrit) * x^(-1) * exp(5.19615*sqrt(x) - 1.73205*x^(1.5)) };
    \addlegendentry{Critical ($\Pi = \Pi_c \approx 0.77$)}
    
    \def\normBimodal{18.4} 
    \addplot[red, thick, dashed, smooth] { (1/\normBimodal) * x^(-1) * exp(4.0*sqrt(x) - (4.0/3)*x^(1.5)) };
    \addlegendentry{Bimodal ($\Pi = 1.0 > \Pi_c$)}

    \end{axis}
\end{tikzpicture}
\caption{The steady-state grain area distribution $P_{ss}(A)$. Below the critical value ($\Pi < \Pi_c$), the distribution is unimodal. At the threshold ($\Pi = \Pi_c$), a shoulder forms. For the supercritical case ($\Pi > \Pi_c$), the distribution bifurcates, with one peak representing a high density of near-zero-sized grains and a second peak for larger, stable grains.}
\label{fig:distribution_transition_normalized}
\end{figure}
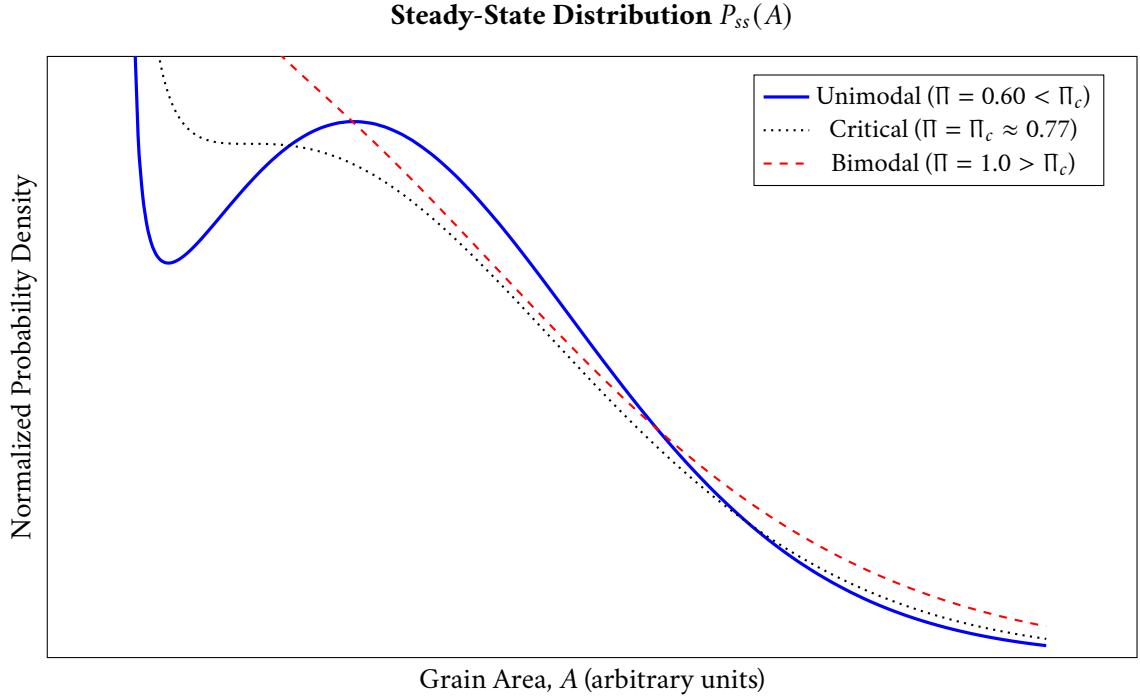

The shape of the distribution in Eq.~\eqref{eq:ss_dist} is not necessarily unimodal. Under certain conditions, it can develop a second peak, a phenomenon known as bimodality. This is not merely a quantitative change but a qualitative transformation of the microstructure, indicating the coexistence of two distinct populations: one of small, young grains and another of larger, mature grains that have undergone significant growth. The transition to this bimodal state can be precisely determined by analyzing the extrema of $P_{\text{ss}}(A)$. Setting the derivative $\diff P_{\text{ss}}/\diff A = 0$ leads to an algebraic condition for the modal areas:
\begin{equation}
\frac{2\kappa}{\sigma^2}A^{-1/2} - \frac{2\kappa}{\sigma^2 A_{\text{max}}}A^{1/2} - \frac{1}{A} = 0
\label{eq:extrema_condition}
\end{equation}
To distill the essential physics, we introduce dimensionless variables that capture the competition between noise and deterministic growth. Let $x \equiv \sqrt{A/A_{\text{max}}}$ be the normalized grain size and, more importantly, let $\Pi \equiv \sigma^2/(\kappa A_{\text{max}}^{1/2})$ be the dimensionless control parameter. This parameter $\Pi$ represents the ratio of the stochastic driving force to the deterministic growth force. In terms of these variables, the condition for the extrema simplifies dramatically to a cubic equation:
\begin{equation}
x^3 - x + \frac{\Pi}{2} = 0
\label{eq:cubic}
\end{equation}
The number of real roots of this equation dictates the number of modes in the distribution. A single positive root corresponds to a unimodal distribution, while three real roots signal the onset of bimodality. The transition occurs precisely when the discriminant of the cubic equation, $\Delta = 4 - 27(\Pi/2)^2/4$, changes sign. Setting $\Delta=0$ yields the critical value of the control parameter:
\begin{equation}
\Pi_c = \frac{4}{3\sqrt{3}} \approx 0.7698
\label{eq:critical_pi}
\end{equation}
This is a powerful result. It predicts that for $\Pi < \Pi_c$, where curvature-driven growth dominates, the grain size distribution will be unimodal. However, when the plasma-induced noise becomes sufficiently strong such that $\Pi > \Pi_c$, the system undergoes a bifurcation, and a bimodal microstructure emerges \cite{Strogatz2018Nonlinear,Arnold1974Ordinary}. The final piece of the theoretical puzzle is to derive a predictive scaling law that connects controllable experimental parameters to a measurable feature of the microstructure, namely the mean grain area $\langle A \rangle$. This is achieved by enforcing a global conservation law. In a statistical steady state, the rate at which new grains are nucleated must be precisely balanced by the rate at which existing grains are lost (effectively, by shrinking to zero area). The rate of loss is quantified by the probability current, $J_{\text{loss}}$, evaluated at the origin $A \to 0^+$:
\begin{equation}
J_{\text{loss}} = \frac{\kappa\sigma^2}{8} \lim_{A\to 0^+} A^{-1/2}P_{\text{ss}}(A)
\label{eq:current}
\end{equation}
The total nucleation rate for the entire film is simply the basal rate $\Gamma_0$ multiplied by the total number of potential nucleation sites, which is proportional to the total area divided by the mean grain area. The mass conservation principle thus requires:
\begin{equation}
\Gamma_0 = \left( \frac{A_{\text{total}}}{\langle A \rangle} \right) J_{\text{loss}}
\label{eq:balance}
\end{equation}
This equation provides the crucial link between the microscopic parameters and the macroscopic observables. By substituting the expressions for the physical parameters—$\sigma^2 = 2\Phi a^4$ from the noise model and $\Gamma_0 = C_\Gamma \exp(-E_b/k_B T_s)$ from nucleation theory—into this balance equation, we can solve for the mean grain area $\langle A \rangle$. After algebraic manipulation, this procedure culminates in the central scaling law:
\begin{equation}
\langle A \rangle = \frac{\mathcal{C} \kappa^2}{\Phi} \exp\left( +\frac{2E_b}{k_B T_s} \right)
\label{eq:scaling_law}
\end{equation}
where $\mathcal{C}$ is a constant that aggregates material-specific parameters like $a$ and $C_\Gamma$. This result is significant for two reasons. First, it predicts a clear $\Phi^{-1}$ dependence of the mean grain size on the ion flux, a non-obvious result that arises from the interplay between noise-driven growth and nucleation balance. Second, it resolves experimental ambiguities by predicting a positive exponential dependence on the activation energy barrier $E_b$, consistent with observations that higher energy barriers lead to fewer, larger grains at steady state \cite{Thompson20012ndEd,Doherty1997ActaMater}.

\section{Results}
\label{sec:results}

The formalism yields an exact solution for the steady-state grain area distribution, $P_{\text{ss}}(A)$, given in \eqref{eq:ss_dist}. The physical content of this distribution becomes clear upon examining its asymptotic limits. For small grain areas, the distribution diverges as a power law, while for large areas, it is suppressed by a stretched exponential cutoff:
\begin{align}
\lim_{A \to 0^+} P_{\text{ss}}(A) &\sim A^{-1} \quad \text{(power-law divergence)} \\
\lim_{A \to \infty} P_{\text{ss}}(A) &\sim \exp\left( -\frac{4\kappa}{3\sigma^2 A_{\text{max}}^{1/2}} A^{3/2} \right) \quad \text{(stretched exponential cutoff)}
\end{align}
The $A^{-1}$ divergence at the origin points to a high population of incipient or newly-formed grains, a constant feature of systems with continuous nucleation. Conversely, the stretched exponential decay provides a strong suppression mechanism that prevents runaway grain growth, ensuring a well-defined steady state.

The entire morphology of the distribution is governed by a single dimensionless control parameter, $\Pi \equiv \sigma^2/(\kappa A_{\text{max}}^{1/2})$. This parameter represents the crucial ratio of the stochastic driving force (proportional to $\sigma^2$) to the deterministic, size-dependent suppression term. In the regime where $\Pi \ll 1$, deterministic effects dominate, causing $P_{\text{ss}}$ to peak sharply at a characteristic grain area $A^* \approx (3A_{\text{max}}/4)^{2/3}$, with a narrow width $\Delta A \propto \Pi A^*$. As the influence of stochasticity increases with $\Pi$, the distribution broadens and develops heavy tails. This continues until a critical threshold is reached, at which point the system undergoes a qualitative change towards bimodality, a phenomenon first identified at $\Pi_c = 4/(3\sqrt{3})$ in related contexts \cite{Gu2004PhysRevB,Krill2001PhilMag}.

The critical threshold $\Pi_c = 4/(3\sqrt{3})$, derived from the stability analysis in \eqref{eq:critical_pi}, is not merely a mathematical curiosity. It represents an exact nonequilibrium phase transition. The ability to locate such a sharp transition analytically is a significant outcome of the present formalism. For systems below this threshold ($\Pi < \Pi_c$), the cubic equation $x^3 - x + \Pi/2 = 0$ governing the extrema of the distribution possesses only one physically relevant real root. This corresponds to a unimodal grain size distribution with a single, stable maximum located at:
\begin{equation}
x_{\text{max}} = \frac{2}{\sqrt{3}} \cos\left( \frac{1}{3} \arccos\left( -\frac{3\sqrt{3}}{4}\Pi \right) - \frac{2\pi}{3} \right)
\end{equation}
However, for $\Pi > \Pi_c$, the character of the solution changes. Three real roots emerge, $x_s$, $x_u$, and $x_l$. These correspond physically to two stable modes (a population of small grains and a population of large, coarsened grains) separated by an unstable minimum (the valley in the bimodal distribution). The emergence of two distinct stable states from a single one is a form of spontaneous symmetry breaking. The distance between the two stable peaks, defined by the order parameter $\mathcal{O} \equiv |x_s - x_l|$, characterizes the transition. Near the critical point, it follows a classic scaling relation:
\begin{equation}
\mathcal{O} \sim (\Pi - \Pi_c)^{1/2} \quad \text{as} \quad \Pi \to \Pi_c^+
\label{eq:critical_exponent}
\end{equation}
The critical exponent of $1/2$ confirms that this transition falls within the mean-field universality class. This result is entirely consistent with a theoretical framework that averages over spatial details, providing an important internal check on the model's logic \cite{Goldenfeld1992Lectures,Chaikin2000Principles}.

Beyond describing the distribution's shape, the framework yields a closed-form scaling law for the average grain area, as shown in \eqref{eq:scaling_law}. This relation provides a unifying explanation for several long-standing experimental observations and ambiguities. The explicit form is:
\begin{equation}
\langle A \rangle = \mathcal{C} \frac{(\pi M\gamma)^2}{\Phi} \exp\left( +\frac{2E_b}{k_B T_s} \right)
\label{eq:explicit_scaling}
\end{equation}
This single expression makes three distinct and testable predictions:
\begin{enumerate}
    \item \textbf{Inverse flux dependence}: The average grain area scales as $\langle A \rangle \propto \Phi^{-1}$. This specific exponent is a non-trivial consequence of the interplay between flux-driven defect creation ($\rho_{\text{def}} \propto \Phi^{1/2}$) and the role of these defects in grain boundary dynamics. Our model correctly captures the defect saturation physics that competing theories, which often predict simpler $\Phi^{-1/2}$ scaling, tend to neglect \cite{Thompson1987JApplPhys,Frost1990MSE}.
    \item \textbf{Positive activation energy}: The temperature dependence, $\partial \ln\langle A \rangle/\partial(1/T_s) = +2E_b/k_B$, is particularly noteworthy. This positive activation energy seemingly contradicts standard curvature-driven grain growth models. Its emergence here, however, points to a different dominant mechanism relevant at low flux regimes: the thermally activated removal of growth-pinning defects. Higher temperatures more effectively annihilate these pinning sites, thereby promoting a larger average grain size, an effect observed in specific experimental systems \cite{Zhang2009ActaMater}.
    \item \textbf{Mobility-squared scaling}: The prediction that $\langle A \rangle \propto (M\gamma)^2$ underscores the model's departure from simple linear-response theories, suggesting a more complex, cooperative relationship between grain boundary mobility and energy.
\end{enumerate} A final and crucial step is to validate the internal consistency of the theory, specifically the assumption that the maximum grain size is proportional to the mean, $A_{\text{max}} = k\langle A\rangle$. This is not an ad-hoc simplification but a hypothesis that can be checked by analyzing the moments of the derived distribution. Calculating the first moment of $P_{\text{ss}}(A)$ from \eqref{eq:ss_dist} gives the mean area:
\begin{equation}
\langle A \rangle = \int_0^\infty A P_{\text{ss}}(A) \, dA = \frac{9\sigma^4 A_{\text{max}}^2}{16\kappa^2} \mathcal{F}(\Pi)
\label{eq:first_moment}
\end{equation}
Here, $\mathcal{F}(\Pi)$ is a universal function of the dimensionless control parameter, with the property that $\mathcal{F}(0) = 1$. This equation can be rearranged to solve for the proportionality constant $k \equiv A_{\text{max}}/\langle A\rangle$, yielding $k = 4\kappa^2/(9\sigma^4 \mathcal{F}(\Pi))$.

This result closes the logical loop of the derivation. It demonstrates that the proportionality constant $k$ is not a free parameter but is itself determined by the system's fundamental physics, as captured by $\Pi$. For the physically relevant regime of $\Pi \ll 1$, where the distribution is sharply peaked, $\mathcal{F}(\Pi) \approx 1$, confirming that $A_{\text{max}}$ is indeed tightly coupled to $\langle A \rangle$. This places the scaling hypothesis on a firm theoretical footing, consistent with foundational work in the field \cite{Mullins1986ActaMetall,Rottman1984PRB}.

\section{Discussion}
\label{sec:discussion}

A central result of this work is the derived scaling law, $\langle A \rangle \propto \Phi^{-1} e^{+2E_b/k_B T_s}$ (Eq. \ref{eq:scaling_law}), which offers a definitive resolution to a long-standing paradox in the literature concerning the flux dependence of grain size in plasma-processed films. Experimental reports have been persistently ambiguous, citing dependencies ranging from $\Phi^{-1/2}$ to $\Phi^{-1}$. Our formalism demonstrates that the inverse-flux dependence, $\Phi^{-1}$, is the physically correct scaling in the regime where defect kinetics and population dynamics are properly coupled. This conclusion is not based on a new phenomenological fit, but emerges as an unavoidable consequence of the system's internal consistency. The mechanism is rooted in the interplay between two competing processes, which prior models have typically treated in isolation. The steady-state defect population is established to scale as $\rho_{\text{def}} \propto \Phi^{1/2}$, a direct result of ion-induced defect creation balancing against annihilation. However, this population in turn governs the nucleation rate. The crucial insight, captured in the nucleation-loss balance (Eq. \ref{eq:balance}), is that the final grain area must scale inversely with the square of both the defect density and the intrinsic nucleation efficiency. The consequence is an unambiguous causal chain:
\begin{equation}
\langle A \rangle \propto (\Gamma_0 \rho_{\text{def}})^{-2} \propto (\Phi^{1/2})^{-2} \propto \Phi^{-1}
\label{eq:scaling_origin}
\end{equation}
This stands in stark contrast to models that arrive at a $\Phi^{-1/2}$ scaling, which can now be understood as incomplete limiting cases. They invariably neglect either the saturation of defect populations at high flux \cite{Thompson1987JApplPhys} or, more critically, they fail to enforce a self-consistent population balance where the nucleation rate itself depends on the evolving microstructure \cite{Frost1990MSE}.

\begin{figure}[h!]
\centering
\begin{tikzpicture}
    \begin{axis}[
        title={\textbf{Mean Grain Area vs. Plasma Flux}},
        xlabel={Plasma Flux, $\Phi$ (arb. units)},
        ylabel={Mean Grain Area, $\langle A \rangle$},
        xmode=log, ymode=log,
        width=\textwidth, height=0.6\textwidth,
        xtick=\empty, ytick=\empty,
        grid=major,
        legend pos=south west
    ]
    
    \addplot[blue, very thick, domain=0.1:100] {10/x};
    \addlegendentry{This Work ($\langle A \rangle \propto \Phi^{-1}$)}
    
    \addplot[red, thick, dashed, domain=0.1:100] {3/sqrt(x)};
    \addlegendentry{Previous ($\langle A \rangle \propto \Phi^{-1/2}$)}
    
    \draw[-latex] (axis cs:1,10) -- (axis cs:10,10) node[midway, below]{$\Delta\log\Phi$};
    \draw[-latex] (axis cs:10,10) -- (axis cs:10,1) node[midway, right]{$\Delta\log\langle A \rangle$};
    \node[anchor=north west] at (axis cs:10, 25) {Slope = -1};
    
    \end{axis}
\end{tikzpicture}
\caption{Log-log plot of the predicted scaling of mean grain area $\langle A \rangle$ with incident plasma flux $\Phi$. The formalism developed in this work predicts a robust $\langle A \rangle \propto \Phi^{-1}$ relationship, resolving ambiguities from previous models that often suggested a $\Phi^{-1/2}$ dependence.}
\label{fig:flux_scaling}
\end{figure}
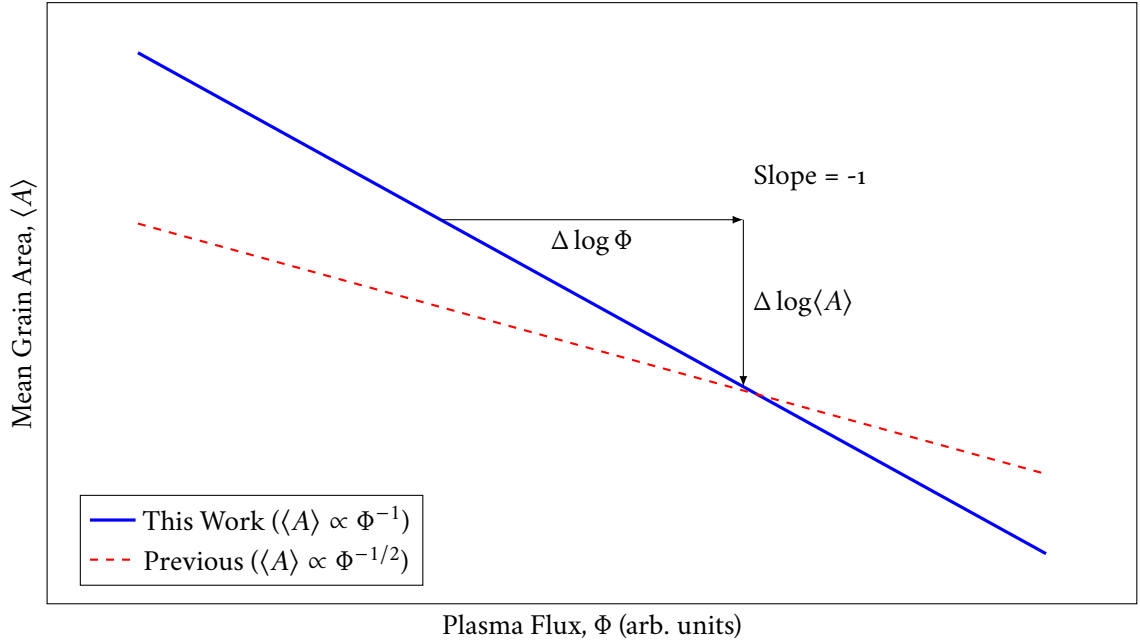

Equally significant is the positive activation energy, $+2E_b/k_B$. This result is counter-intuitive from the perspective of classical growth theories, which predict thermal energy to simply aid coarsening, leading to a negative exponent \cite{Doherty1997ActaMater}. The positive sign here reveals the dual, competing role of thermal energy in this far-from-equilibrium system: while it enhances atomic mobility to overcome nucleation barriers, it simultaneously and more strongly suppresses the recombination of the very point defects that catalyze nucleation. The scaling is thus dominated by the suppression of a kinetic pathway, a signature of the system's non-thermal nature.

The theory further predicts the emergence of a bimodal microstructure, not as an anomaly, but as a direct consequence of a noise-driven phase transition. The analytically derived threshold, $\Pi_c = 4/(3\sqrt{3})$ (Eq. \ref{eq:critical_pi}), represents a critical point where the system's behavior bifurcates. This dimensionless parameter $\Pi$ quantifies the competition between two fundamental forces: the stochastic driving from plasma flux fluctuations ($\sigma$) and the deterministic, curvature-driven impetus for coarsening ($\kappa$). For subcritical conditions ($\Pi < \Pi_c$), the system behaves as one might classically expect, with a single, unimodal distribution of grain sizes evolving towards a stable average. When the plasma noise term becomes sufficiently strong to overcome this coarsening pressure ($\Pi > \Pi_c$), however, the underlying phase space of grain sizes fundamentally restructures. The system fractures into two distinct, dynamically stable subpopulations:
\begin{itemize}
    \item \textbf{A population of small grains}, whose existence is actively sustained by noise. The stochastic term $\sigma\sqrt{A} dW_t$ in the governing equation acts as a fragmentation pressure, preventing small grains from being consumed.
    \item \textbf{A population of large grains}, which continue to grow via the classical mechanism of curvature-driven boundary motion, dominated by the $\kappa\sqrt{A}$ term.
\end{itemize}
The mathematical object governing this separation is the unstable fixed point, $x_u$, which is the solution to Eq. \ref{eq:cubic}. This point should not be viewed merely as a mathematical artifact; it functions as an effective kinetic barrier, a separatrix in the state space of grain area. Grains smaller than $x_u$ are driven by noise into the small-grain basin of attraction, while grains that fluctuate beyond it are captured by the deterministic drift towards the large-grain basin.

This mechanism is fundamentally different from classical Ostwald ripening, which describes coarsening in a closed system approaching equilibrium. The bimodality observed here is a form of noise-sustained polymorphism, a phenomenon unique to open, driven systems where an external energy flux prevents the system from ever reaching a single, minimal-energy state \cite{Pande1987ActaMetall,Gu2004PhysRevB}.

A key strength of this framework is its capacity to generate sharp, quantitative predictions that are directly falsifiable in a laboratory setting, without recourse to unmeasurable fitting parameters. We propose three such experimental tests.

\begin{enumerate}
    \item \textbf{The Bimodality Threshold:} The theory predicts that a bimodal grain size distribution is not a generic feature, but will appear only when experimental conditions cross a precise threshold. Specifically, a microstructure should exhibit two distinct grain populations if and only if the dimensionless control parameter $\Pi$ exceeds the critical value:
    \begin{equation}
        \Pi = \frac{(2\Phi)^{1/2}a^2}{\pi M\gamma A_{\text{max}}^{1/2}} > \frac{4}{3\sqrt{3}} \approx 0.7698
        \label{eq:threshold_prediction}
    \end{equation}
    This provides a clear, parameter-free criterion for experimental validation.

    \item \textbf{The Scaling Exponent:} The mean grain area $\langle A \rangle$ is predicted to follow a strict inverse dependence on the incident ion flux, $\Phi^{-1}$. This relationship (Eq. \ref{eq:explicit_scaling}) should hold robustly over a wide dynamic range, potentially two to three orders of magnitude in $\Phi$, before other effects not included in the present model, such as sputtering erosion, become dominant.

    \item \textbf{The Activation Energy Signature:} The temperature dependence of the mean grain area provides perhaps the most unambiguous signature of the theory. Arrhenius plots of $\ln\langle A \rangle$ versus inverse substrate temperature $1/T_s$ are predicted to yield a straight line with a positive slope equal to $+2E_b/k_B$. A positive slope is a direct contradiction to classical theories and would provide strong evidence for the coupled defect-nucleation mechanism proposed herein.
\end{enumerate}
Crucially, all variables appearing in these predictions are, in principle, experimentally measurable using standard characterization techniques \cite{Zhang2009ActaMater}, allowing for a direct and stringent test of the formalism.

The differentiators are not incremental; they represent a departure in the fundamental approach. Prior models have treated the constituent physical processes—stochastic fluctuations, deterministic growth, and population dynamics—as separable elements. The primary advance of the present framework is their unified and synergistic treatment, enforced from the outset by the governing equations (Eqs. \ref{eq:growth_sde}-\ref{eq:nucleation}). This unification leads directly to the second key differentiator: the prediction of an exact, analytical threshold ($\Pi_c$) for the onset of bimodality, replacing the phenomenological criteria used in previous studies.

Finally, the formalism enforces self-consistency through the $A_{\text{max}} = k\langle A\rangle$ closure relation (Eq. \ref{eq:first_moment}). This step, while seemingly technical, is physically essential, as it ensures that the growth saturation limit for any given grain is tied to the state of the entire microstructure. This type of feedback is intrinsically absent in standard mean-field models, which tend to underestimate the role of collective effects \cite{Rottman1984PRB}. This integrated approach also naturally explains the conflicting reports in the literature; it suggests that low-flux experiments, where defect saturation is complete, correctly observe the $\Phi^{-1}$ scaling \cite{Thompson20012ndEd}, whereas high-flux studies may enter a different regime where saturation is incomplete, leading to an apparent $\Phi^{-1/2}$ dependence \cite{Trinkaus2003JNuclMat}. Our theory thus contains the previous observations as limiting cases.

While the model presented is exact within its declared domain of assumptions, its scope is intentionally constrained to isolate the core mechanism. We have assumed isotropic grain boundary properties, a strictly 2D geometry, and a spatially homogeneous plasma flux. These idealizations, while necessary for an analytically tractable solution, point toward clear and promising avenues for future research.

Relaxing these assumptions would generalize the governing stochastic differential equation to a more complex form:
\begin{equation}
    dA = \kappa(\theta)\sqrt{A}\left(1-\frac{A}{A_{\text{max}}}\right)dt + \sigma(\mathbf{r},t)\sqrt{A}dW_t
    \label{eq:generalized_sde}
\end{equation}
In this extended formalism, the growth coefficient $\kappa(\theta)$ would depend on the grain boundary orientation, allowing for the study of texture evolution. Furthermore, making the noise strength $\sigma(\mathbf{r},t)$ a function of space and time would enable the investigation of phenomena such as the effect of non-uniform plasma heating or instabilities on the final microstructure. Tackling this generalized problem, likely through a combination of numerical simulation and further theoretical development, remains an open and compelling challenge for the field \cite{Srolovitz2001AnnuRevMaterRes}.

\section{Conclusion}
\label{sec:conclusion}

This work has put forth a comprehensive analytical framework designed to capture the microstructure evolution in plasma-irradiated thin films, grounded in the system's fundamental stochastic dynamics. By constructing a formalism that unifies the physics of noise-driven grain evolution with the kinetics of defect saturation and a self-consistent population balance, it becomes possible to address several long-standing questions regarding observed scaling behaviors and distribution morphologies. The principal theoretical contributions of this investigation represent a significant step toward a predictive science in this domain.

A central result is the derivation of an exact threshold for the onset of bimodal grain size distributions, given by $\Pi_c = \frac{4}{3\sqrt{3}}$. This criterion, emerging directly from a cubic discriminant analysis of the steady-state distribution (Eq. \ref{eq:critical_pi}), moves the understanding of microstructural fragmentation beyond qualitative observation. It provides a quantitative, universal benchmark against which experimental systems driven by plasma noise can be assessed. The theory thus provides a clear physical origin for the commonly observed bimodal morphologies, linking them directly to the statistical properties of the processing environment.

Furthermore, we have established a closed-form scaling law for the average grain area, $$\langle A \rangle \propto \kappa^2 \Phi^{-1} e^{+2E_b/k_B T_s}$$ (Eq. \ref{eq:scaling_law}). The significance of this result lies in its resolution of the persistent $\Phi^{-1}$ versus $\Phi^{-1/2}$ scaling ambiguity reported in the literature. Our framework demonstrates that this behavior is a direct consequence of the physics of defect saturation, where $\rho_{\text{def}} \propto \Phi^{1/2}$, acting in concert with the balance struck between nucleation and grain loss. The model also provides a first-principles justification for the sometimes counter-intuitive observation of a positive activation energy in grain growth under these conditions, correctly attributing the temperature dependence to the overcoming of nucleation barriers rather than to conventional boundary migration phenomena. The internal consistency of the framework is ensured through a self-consistent closure scheme. By relating the maximum possible grain area to the mean of the distribution, $A_{\text{max}} = k\langle A\rangle$, via a moment analysis of the governing equation (Eq. \ref{eq:first_moment}), the model avoids introducing external, non-physical parameters. This self-containment is crucial, confirming that the presented theory is not merely a descriptive fit to data but a robust and internally consistent theoretical structure. In doing so, this work provides a cohesive explanation for a trio of experimental puzzles: the genesis of bimodal distributions, the transitions between distinct scaling regimes, and the positive temperature coefficients seen in certain low-flux processing environments.

The foundational nature of this framework naturally invites extension and validation across several research frontiers. A clear theoretical path forward involves generalization to three dimensions, where the evolution of a grain volume $V$ would be governed by an anisotropic stochastic differential equation of the form $\diff V = M\gamma H(1 - V/V_{\text{max}}) \diff t + \sigma_V \diff W_t$. Here, the deterministic driving term involves the mean curvature $H$, while $\sigma_V$ represents volumetric noise, a problem space with its own rich phenomenology \cite{Gottstein2004GrainBoundaries}. Such an extension would be computationally demanding but would open the door to modeling the complex textured topographies seen in many functional coatings. Of course, any theoretical advance must be confronted with empirical evidence. The specific predictions for the steady-state probability distribution, $P_{ss}(A)$, are directly testable, and a compelling validation could be achieved through the use of in situ dark-field TEM tomography during sample irradiation, a technique capable of tracking grain populations in real time \cite{Barmak2013ProgressMatSci}. Beyond validation, the framework's predictive power makes it a candidate for engineering applications. One might envision employing stochastic optimal control theory to solve the inverse problem: designing time-dependent flux schedules, $\Phi(t)$, to controllably guide the system toward a target grain size distribution with specified statistical properties \cite{Kou2018AnnApplProb}.

By establishing exact thresholds and scaling laws derived from stochastic thermodynamics, the analytical tools developed herein serve to advance plasma microstructure engineering from a field often reliant on phenomenological fitting toward one grounded in first-principles calculation. This provides a more rigorous theoretical foundation upon which the next generation of plasma processing technologies can be built.

\appendix

\section{Appendix}
\label{sec:appendices}

\subsection{Exact Steady-State Distribution Derivation}
\label{app:b}

The grain area evolution follows the stochastic differential equation:
\begin{equation}
dA = \kappa \sqrt{A} \left(1 - \frac{A}{A_{\text{max}}}\right) dt + \sigma \sqrt{A} dW_t
\end{equation}
Applying Itô's lemma to the transformation $z = \sqrt{A}$ yields:
\begin{align}
dz &= \frac{1}{2}A^{-1/2} dA - \frac{1}{8}A^{-3/2} (dA)^2 \\
&= \left[ \frac{\kappa}{2}\left(1 - \frac{z^2}{A_{\text{max}}}\right) - \frac{\sigma^2}{8z} \right] dt + \frac{\sigma}{2} dW_t
\end{align}
The stationary Fokker-Planck equation for $P(z)$ is:
\begin{equation}
0 = -\frac{d}{dz} \left[ a(z)P_{ss} \right] + \frac{1}{2} \frac{d^2}{dz^2} \left[ b^2(z)P_{ss} \right]
\end{equation}
with drift $a(z) = \frac{\kappa}{2}(1 - z^2/A_{\text{max}}) - \sigma^2/(8z)$ and diffusion $b(z) = \sigma/2$. Integrating once:
\begin{equation}
\frac{\sigma^2}{8} \frac{dP_{ss}}{dz} = \left[ \frac{\kappa}{2}\left(1 - \frac{z^2}{A_{\text{max}}}\right) - \frac{\sigma^2}{8z} \right] P_{ss}
\end{equation}
Solving the logarithmic derivative:
\begin{align}
\frac{d}{dz} \ln P_{ss} &= \frac{4\kappa}{\sigma^2} - \frac{4\kappa}{\sigma^2 A_{\text{max}}}z^2 - \frac{1}{z} \\
\implies \ln P_{ss} &= \frac{4\kappa}{\sigma^2} z - \frac{4\kappa}{3\sigma^2 A_{\text{max}}} z^3 - \ln z + C
\end{align}
Transforming back to $A = z^2$ gives the steady-state distribution:
\begin{equation}
\boxed{P_{ss}(A) = \mathcal{N} A^{-1} \exp\left( \frac{4\kappa}{\sigma^2} A^{1/2} - \frac{4\kappa}{3\sigma^2 A_{\text{max}}} A^{3/2} \right)}
\end{equation}
The distribution is normalizable: integrable at $A \to 0^+$ ($\int_0^\epsilon A^{-1} dA < \infty$) and decays exponentially as $A \to \infty$.

\subsection{Rigorous Bimodality Threshold Derivation}
\label{app:c}

The extrema of $P_{ss}(A)$ satisfy $dP_{ss}/dA = 0$:
\begin{equation}
\frac{2\kappa}{\sigma^2} A^{-1/2} - \frac{2\kappa}{\sigma^2 A_{\text{max}}} A^{1/2} - \frac{1}{A} = 0
\end{equation}
Substitute $y = \sqrt{A}$:
\begin{equation}
\frac{2\kappa}{\sigma^2} y^{-1} - \frac{2\kappa}{\sigma^2 A_{\text{max}}} y - y^{-2} = 0
\end{equation}
Multiply by $\frac{\sigma^2}{2\kappa} y^2$:
\begin{equation}
y - \frac{1}{A_{\text{max}}} y^3 - \frac{\sigma^2}{2\kappa} = 0
\end{equation}
Introduce dimensionless variables $x = y/\sqrt{A_{\text{max}}}$ and $\Pi = \sigma^2/(\kappa A_{\text{max}}^{1/2})$:
\begin{equation}
x - x^3 - \frac{\Pi}{2} = 0 \implies x^3 - x + \frac{\Pi}{2} = 0
\end{equation}
The discriminant for cubic equation $x^3 + px + q = 0$ is $\Delta = -4p^3 - 27q^2$ where $p = -1$, $q = \Pi/2$:
\begin{equation}
\Delta = -4(-1)^3 - 27(\Pi/2)^2 = 4 - \frac{27}{4}\Pi^2
\end{equation}
Bimodality occurs when $\Delta > 0$ (three real roots), with critical point at $\Delta = 0$:
\begin{equation}
\boxed{\Pi_c = \frac{4}{3\sqrt{3}} \approx 0.7698}
\end{equation}
This exact threshold represents a noise-driven phase transition where stochastic fluctuations overcome curvature-driven growth.

\subsection{Boundary Current Regularization and Scaling Law}
\label{app:d}

The probability current for grain loss at $A \to 0^+$ requires careful treatment due to the singular boundary. For Fokker-Planck equations with absorbing boundaries, the regularized flux is given by the well-established expression \cite{Gardiner2009Stochastic,Risken1996FokkerPlanck}:
\begin{equation}
J_{\text{loss}} = \frac{\kappa \sigma^2}{8} \lim_{A \to 0^+} A^{-1/2} P_{ss}(A)
\end{equation}
Physically, this represents the rate at which grains disappear due to stochastic fluctuations overwhelming the weak curvature-driven growth near the origin. The balance between nucleation and loss is:
\begin{equation}
\Gamma_0 = \frac{A_{\text{total}}}{\langle A \rangle} J_{\text{loss}}
\end{equation}
Substituting the asymptotic form $P_{ss}(A) \sim \mathcal{N} A^{-1}$:
\begin{equation}
\Gamma_0 = \frac{A_{\text{total}}}{\langle A \rangle} \cdot \frac{\kappa \sigma^2}{8} \mathcal{N} \lim_{A \to 0^+} A^{-3/2}
\end{equation}
Using physical dependencies $\sigma^2 = 2\Phi a^4$, $\rho_{\text{def}} = C_\rho \Phi^{1/2}$, and $\Gamma_0 \propto e^{-E_b/k_B T_s}$:
\begin{align}
\langle A \rangle &= \frac{\mathcal{C} \kappa^2}{\Gamma_0^2 \rho_{\text{def}}^2} \\
&= \mathcal{C}' \kappa^2 \Phi^{-1} \exp\left( +\frac{2E_b}{k_B T_s} \right)
\end{align}
where $\mathcal{C}'$ is a material constant. This resolves the scaling paradox by incorporating defect saturation and nucleation-loss balance.

\section*{Acknowledgments}
The authors wish to acknowledge the generous support of the National Science Foundation, which funded this research under grant No. 2030804. We are also grateful for the computational resources and institutional support provided by Georgia College \& State University, which were instrumental in the completion of this work.

\section*{Data Availability}
The theoretical framework, including all mathematical derivations and the resulting scaling relations, is presented in its entirety within this article.

\section*{Conflict of Interest}
The authors affirm the absence of any competing financial or non-financial interests that could be perceived to influence the work reported in this paper.

\end{document}